\documentclass[aps,prl,twocolumn,superscriptaddress,floatfix,showpacs]{revtex4}
\usepackage[dvipdfmx]{graphicx}
\usepackage[dvipdfmx]{color}
\usepackage{amsmath}
\usepackage{amssymb}
\usepackage{dcolumn}
\usepackage{bm}
\usepackage{latexsym} 
\usepackage{setspace}
\usepackage{graphicx}
\usepackage{color}
\usepackage{here}
\begin{document}
\title{Magnon Polarons in the Spin Seebeck Effect}
\author{Takashi Kikkawa}
\email{t.kikkawa@imr.tohoku.ac.jp}
\affiliation{Institute for Materials Research, Tohoku University, Sendai 980-8577, Japan}
\affiliation{WPI Advanced Institute for Materials Research, Tohoku University, Sendai
980-8577, Japan}
\author{Ka Shen}
\affiliation{Kavli Institute of NanoScience, Delft University of Technology, Lorentzweg
1, 2628 CJ Delft, The Netherlands}
\author{Benedetta Flebus}
\affiliation{Institute for Theoretical Physics and Center for Extreme Matter and Emergent
Phenomena, Utrecht University, Leuvenlaan 4, 3584 CE Utrecht, The Netherlands}
\author{Rembert A. Duine}
\affiliation{Institute for Theoretical Physics and Center for Extreme Matter and Emergent
Phenomena, Utrecht University, Leuvenlaan 4, 3584 CE Utrecht, The Netherlands}
\affiliation{Department of Applied Physics, Eindhoven University of Technology, P.O. Box
513, 5600 MB Eindhoven, The Netherlands}
\author{Ken-ichi Uchida}
\email[Present address: National Institute for Materials Science, Tsukuba 305-0047, Japan.]{}
\affiliation{Institute for Materials Research, Tohoku University, Sendai 980-8577, Japan}
\affiliation{PRESTO, Japan Science and Technology Agency, Saitama 332-0012, Japan}
\affiliation{Center for Spintronics Research Network, Tohoku University, Sendai 980-8577,
Japan}
\author{Zhiyong Qiu}
\affiliation{WPI Advanced Institute for Materials Research, Tohoku University, Sendai
980-8577, Japan}
\affiliation{Spin Quantum Rectification Project, ERATO, Japan Science and Technology
Agency, Sendai 980-8577, Japan}
\author{Gerrit E. W. Bauer}
\affiliation{Institute for Materials Research, Tohoku University, Sendai 980-8577, Japan}
\affiliation{WPI Advanced Institute for Materials Research, Tohoku University, Sendai
980-8577, Japan}
\affiliation{Kavli Institute of NanoScience, Delft University of Technology, Lorentzweg
1, 2628 CJ Delft, The Netherlands}
\affiliation{Center for Spintronics Research Network, Tohoku University, Sendai 980-8577,
Japan}
\author{Eiji Saitoh}
\affiliation{Institute for Materials Research, Tohoku University, Sendai 980-8577, Japan}
\affiliation{WPI Advanced Institute for Materials Research, Tohoku University, Sendai
980-8577, Japan}
\affiliation{Center for Spintronics Research Network, Tohoku University, Sendai 980-8577,
Japan}
\affiliation{Spin Quantum Rectification Project, ERATO, Japan Science and Technology
Agency, Sendai 980-8577, Japan}
\affiliation{Advanced Science Research Center, Japan Atomic Energy Agency, Tokai
319-1195, Japan}
\date{\today}
\begin{abstract}
Sharp structures in the magnetic field-dependent spin Seebeck effect (SSE) voltages of Pt/Y$_{3}$Fe$_{5}$O$_{12}$ at low temperatures are attributed to the magnon-phonon interaction. Experimental results are well reproduced by a Boltzmann theory that includes magnetoelastic coupling. The SSE anomalies coincide with magnetic fields tuned to the threshold of magnon-polaron formation. The effect gives insight into the relative quality of the lattice and magnetization dynamics.
\end{abstract}
%
\maketitle
\begin{figure}[H]
\begin{center}
\includegraphics{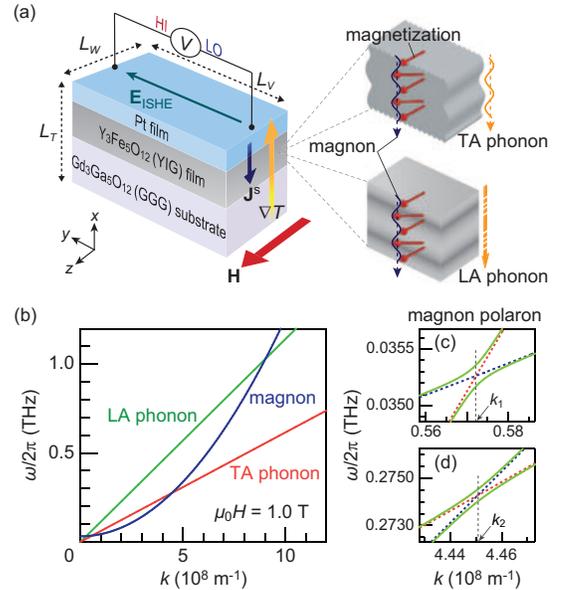}
\end{center}
\caption{(a) The longitudinal SSE in the Pt/YIG/GGG sample, where $\mathbf{E}_{\mathrm{ISHE}}$ denotes the electric
field induced by the ISHE. The close-up of the upper (lower) right shows a
schematic illustration of a propagating magnon and TA (LA) phonon. 
(b) Magnon [Eq. (\protect\ref%
{equ:magnon}) with $\protect\mu _{0}M_{\mathrm{s}}=0.2439~%
\text{T}$, $\protect\mu _{0}H=1.0~\text{T}$, and $\protect\theta =\protect%
\pi /2$], TA-phonon ($\protect\omega =c_{\perp }k$), and LA-phonon ($\protect%
\omega =c_{||}k$) dispersion relations for the parameters in Table I.
(c),(d) Magnon polarons at the (anti)crossings between the magnon and
TA-phonon branches at (c) lower and (d) higher wave numbers, where $\mathbf{k%
}\parallel \hat{\mathbf{x}}$ ($\protect\theta =\protect\pi /2$ and $\protect%
\phi =0$) and $\mathbf{H}\parallel \hat{\mathbf{z}}$. }
\label{fig:1}
\end{figure}

%
%
%
The spin Seebeck effect (SSE) \cite%
{SSE_first-LSSE,SSE_Xiao2010PRB,SSE_Adachi2011PRB,Zhang-Zhang2012PRB,SSE_Hoffman2013PRB,SSE_Rezende2014PRB,SSE_Wu2015PRL,SSE_Kikkawa2015PRB,SSE_Kehlberger2015PRL,SSE_Jin2015PRB,SSE_Ritzmann2015PRB,SSE_Guo2015arXiv,SSE_Aqeel2015PRB,SSE_Seki2015PRL,SSE_Wu2016PRL,SSE_GdIG,SSE_Rezende2016JMMM,LSSE_magnon-drag1,SSE_review}
refers to the generation of a spin current (${\mathbf{J}}^{s}$) as a result
of a temperature gradient ($\nabla T$) in magnetic materials. It is well
established for magnetic insulators with metallic contacts, at which a
magnon flow is converted into a conduction-electron spin current by the
interfacial exchange interaction \cite{Tserkovnyak05} and detected as a
transverse electric voltage via the inverse spin Hall effect (ISHE) \cite%
{ISHE_Azevedo,ISHE_Saitoh,ISHE_Costache,ISHE_Valenzuela,Weiler2012PRL,Bai2015PRL,ISHE_Sinova}
[see Fig.~\ref{fig:1}(a)]. The SSE provides a sensitive probe for spin
correlations in magnetic materials \cite%
{SSE_Kikkawa2015PRB,SSE_Jin2015PRB,SSE_Aqeel2015PRB,SSE_Seki2015PRL,SSE_Wu2016PRL,SSE_GdIG}%
.\par
The ferrimagnetic insulator yttrium-iron-garnet Y$_{3}$Fe$_{5}$O$_{12}$
(YIG) is ideal for SSE measurements \cite{SSE_review}, exhibiting a long
magnon-propagation length \cite{Boona2014PRB,Giles2015PRB,Cornelissen2016PRB}%
, high Curie temperature ($\sim 560~\text{K}$) \cite{YIG_Gilleo-Geller}, and
high resistivity owing to a large band gap ($\sim 2.9~\text{eV}$) \cite%
{Metselaar}. The magnon and phonon dispersion relations in YIG are well
known \cite%
{Harris1963PhysRev,Keffer-text,YIG_saga,YIG_Srivastava,Strauss-text,Gurevich-Melkov_text}%
. The magnon dispersion in the relevant regime reads 
\begin{equation}
\omega _{\mathbf{k}}=\sqrt{D_{\mathrm{ex}}k^{2}+\gamma \mu _{0}H}\sqrt{D_{%
\mathrm{ex}}k^{2}+\gamma \mu _{0}H+\gamma \mu _{0}M_{\mathrm{s}}\mathrm{%
sin}^{2}\theta },  \label{equ:magnon}
\end{equation}%
where $\omega $, $\mathbf{k}$, $\theta ,\gamma $, and $\mu _{0}M_{\mathrm{s}%
}$, are the angular frequency, wave vector $\mathbf{k}$ with length $k$,
angle $\theta $ with the external magnetic field $\mathbf{H}$ (of magnitude $%
H$), gyromagnetic ratio, and saturation magnetization, respectively 
\cite{Harris1963PhysRev,Keffer-text,YIG_saga,YIG_Srivastava}. The exchange
stiffness coefficient $D_{\mathrm{ex}}$ as well as the transverse-acoustic (TA)
and longitudinal-acoustic (LA) sound velocities for YIG are summarized in
Table \ref{tab:1} and the dispersion relations are plotted in Fig. \ref%
{fig:1}(b). \par
In this Letter, we report the observation of a resonant enhancement of the
SSE. The experimental results are well reproduced by a theory for the ~thermally ~induced
\begin{figure*}[tbh]
\begin{center}
\includegraphics[width=15.5cm]{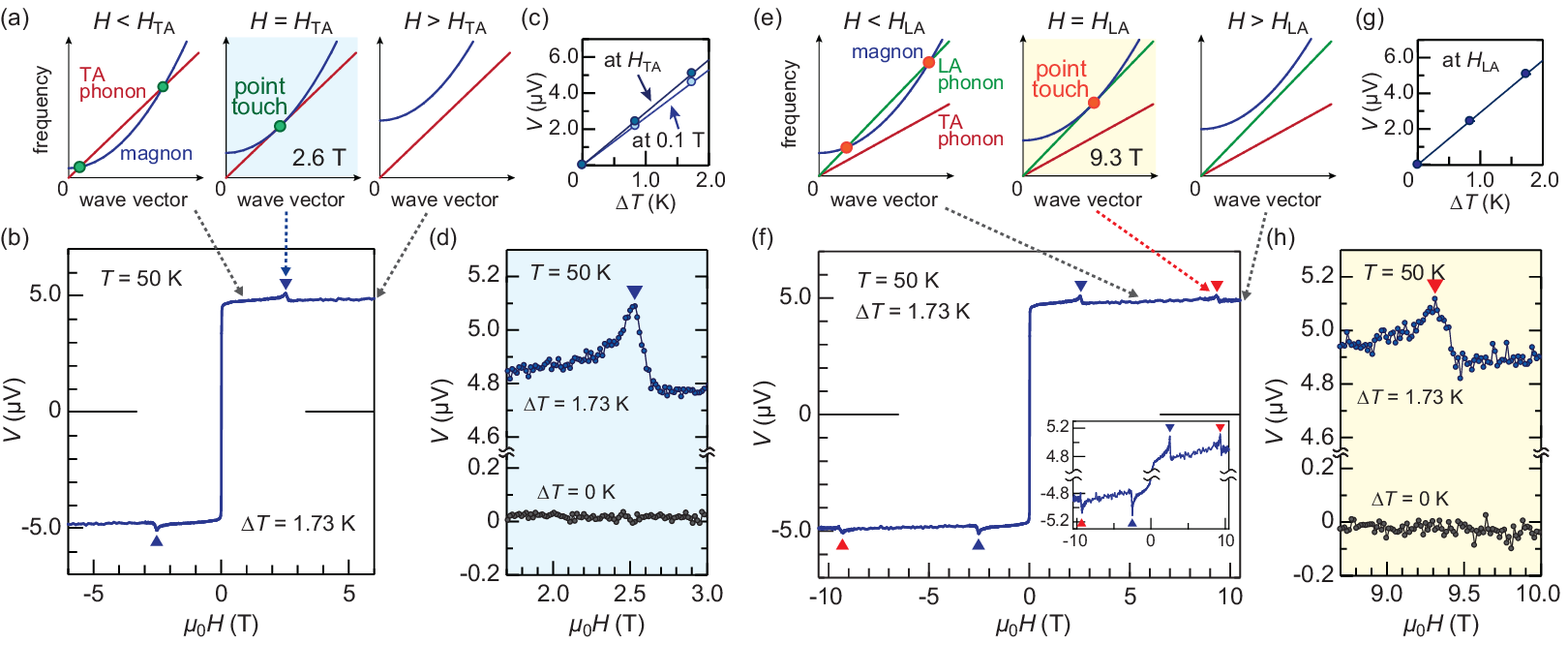}
\end{center}
\caption{(a) Magnon and TA-phonon dispersion relations for YIG when $H<H_{%
\mathrm{TA}}$, $H=H_{\mathrm{TA}}$, and $H>H_{\mathrm{TA}}$. (b) $V\left(
H\right) $ of the Pt/YIG/GGG sample for $%
\Delta T=1.73~\text{K}$ at $T=50~\text{K}$ for $\left\vert \mu_{0} H\right\vert <6.0~%
\text{T}$. (c) $V\left( \Delta T\right) $ of the Pt/YIG/GGG sample at $\mu_{0} H=0.1~\text{T}$ and $\mu_{0}H_{\mathrm{TA}}$. (d)
Magnified view of $V\left( H\right) $ around $H_{\mathrm{TA}}$. (e) Magnon,
TA-phonon, and LA-phonon dispersion relations for YIG when $H<H_{\mathrm{LA}}
$, $H=H_{\mathrm{LA}}$, and $H>H_{\mathrm{LA}}$. (f) $V\left( H\right) $ of
the Pt/YIG/GGG sample for $\Delta T=1.73~%
\text{K}$ at $T=50~\text{K}$ for $\left\vert \mu_{0} H\right\vert <10.5~\text{T}$.
The inset to (f) is a magnified view of $V\left( H\right) $ for $4.6~\mathrm{%
\protect\mu }\text{V}<\left\vert V\right\vert <5.3~\mathrm{\protect\mu }%
\text{V}$. (g) $V\left( \Delta T\right) $ of the Pt/YIG/GGG sample at $H=H_{\mathrm{LA}}$. (h) Magnified view of $%
V\left( H\right) $ around $H_{\mathrm{LA}}$. The $V$ peaks at $H_{%
\mathrm{TA}}$ and $H_{\mathrm{LA}}$ are marked by blue and red triangles,
respectively.}
\label{fig:2}
\end{figure*}
\begin{table}[b]
\caption{Parameters for the magnon and phonon dispersion relations of YIG 
\protect\cite%
{Keffer-text,YIG_Srivastava,YIG_saga,Strauss-text,Gurevich-Melkov_text,Shinozaki1961PhysRev}%
.}\setlength{\belowcaptionskip}{0mm}
\par
\begin{center}
\begin{tabular}{llll}
\hline\hline
& ~Symbol~ & ~~Value & ~Unit \\ \hline
Exchange stiffness & ~~~~$D_{\mathrm{ex}}$ & $7.7\times 10^{-6}$ & ~$%
\text{m}^2/\text{s}$ \\ 
TA-phonon sound velocity & ~~~~~$c_{\perp }$ & $3.9\times 10^{3}$ & ~$\text{m%
}/\text{s}$ \\ 
LA-phonon sound velocity & ~~~~~$c_{||}$ & $7.2\times 10^{3}$ & ~$\text{m}/%
\text{s}$ \\ \hline\hline
\end{tabular}%
\end{center}
\label{tab:1}
\end{table}
\noindent magnon flow in which
the magnetoelastic interaction is taken into
account. We interpret the experiments as evidence for a strong
magnon-phonon coupling at the crossings between the magnon and phonon
dispersion curves, i.e., the formation of hybridized excitations called magnon polarons \cite{Kamra2015PRB,Shen2015PRL}. \par
The sample is a 5-nm-thick Pt film sputtered on the (111) surface of a 4-$%
\mathrm{\mu }\text{m}$-thick single-crystalline YIG film grown on a
single-crystalline Gd$_{3}$Ga$_{5}$O$_{12}$ (GGG) (111) substrate by liquid
phase epitaxy \cite{Qiu2013APL}. The sample was then cut into a rectangular
shape with $L_{V}=4.0~\text{mm}$ (length), $L_{W}=2.0~\text{mm}$ (width),
and $L_{T}=0.5~\text{mm}$ (thickness). SSE measurements were carried out in
a longitudinal configuration \cite{SSE_first-LSSE,SSE_review} [see Fig. \ref%
{fig:1}(a)], where the temperature gradient $\nabla T$ is applied normal to
the interfaces by sandwiching the sample between two sapphire plates, on top
of the Pt layer (at the bottom of the GGG substrate) stabilized to $T_{%
\mathrm{H}}$ ($T_{\mathrm{L}}$) with a temperature difference $\Delta T=T_{%
\mathrm{H}}-T_{\mathrm{L}}$ $(>0)$. $\Delta T$ was measured with two
calibrated Cernox thermometers. A uniform magnetic field $\mathbf{H=}H%
\mathbf{\hat{z}}$ was applied by a superconducting solenoid magnet. We
measured the dc electric voltage difference $V$ between the ends of the Pt
layer with a highly resolved field scan, i.e., at intervals of $15~%
\text{mT}$ and waiting for $\sim 30~\text{sec}$ after each step.\par 
Figure \ref{fig:2}(b) shows the measured $V\left( H\right) $ of the Pt/YIG sample at $T=50~\text{K%
}$. A clear signal appears by applying the temperature difference $\Delta T$ and its sign is reversed when reversing the magnetization. The magnitude
of $V$ at $\mu_{0} H=0.1~\text{T}$ is proportional to $\Delta T$ [see Fig. \ref%
{fig:2}(c)]. These results confirm that $V$ is generated by the SSE \cite%
{SSE_review}. \par 
Owing to the high resolution of $H$, we were able to resolve a fine peak
structure at $\mu_{0}H\sim 2.6~\text{T}$ that is fully reproducible. A magnified
view of the $V$-$H$ curve is shown in Fig. \ref{fig:2}(d), where the anomaly
is marked by a blue triangle. Since the structures scale with $\Delta T$
[see Figs. \ref{fig:2}(c) and \ref{fig:2}(d)], they must stem from the SSE.\par
The peak appears for the field $H_{\mathrm{TA}}$ at which according to the
parameters in Table \ref{tab:1} the magnon dispersion curve touches the TA-phonon dispersion curve. By increasing $H$, the magnon dispersion shifts
toward high frequencies due to the Zeeman interaction ($\propto \gamma \mu
_{0}H$), while the phonon dispersion does not move. At $\mu_{0}H=0$, the magnon
branch intersects the TA-phonon curve twice [see Fig. \ref%
{fig:2}(a)]. With increasing $H$, the TA-phonon branch becomes tangential to the
magnon dispersion at $\mu_{0}H=2.6~\text{T}$ and detaches at higher fields [see Fig. \ref{fig:2}(a)]. If the anomaly is indeed linked to the \textquotedblleft
touch\textquotedblright\ condition, there should be another peak associated with the LA-phonon branch. Based on the parameters in Table \ref{tab:1}, we
evaluated the magnon$-$LA-phonon touch condition at $\mu_{0}H_{\mathrm{LA}}\sim
9.3~\text{T.}$ We then upgraded the equipment with a stronger magnet and
subsequently investigated the high-field dependence of the SSE. \par
Figure \ref{fig:2}(f) shows the dependence $V\left( H\right) $ of the Pt%
/YIG sample at $T=50~\text{K}$, measured between $\mu_{0}H=\pm
10.5~\text{T}$. Indeed, another peak appeared at $\mu_{0}H_{\mathrm{LA}}\sim 9.3~%
\text{T}$ precisely at the estimated field value at which the LA-phonon
branch touches the magnon dispersion [see Fig. \ref{fig:2}(e)], sharing the
characteristic features of the SSE; i.e., it appears only when $\Delta T\neq 0$
and exhibits a linear-$\Delta T$ dependence [see Figs. \ref{fig:2}(g) and %
\ref{fig:2}(h)]. For $\mu_{0}H>9.3~\text{T}$ the $V$-$H$ curves remain smooth. \par
\begin{figure}[tbh]
\begin{center}
\includegraphics{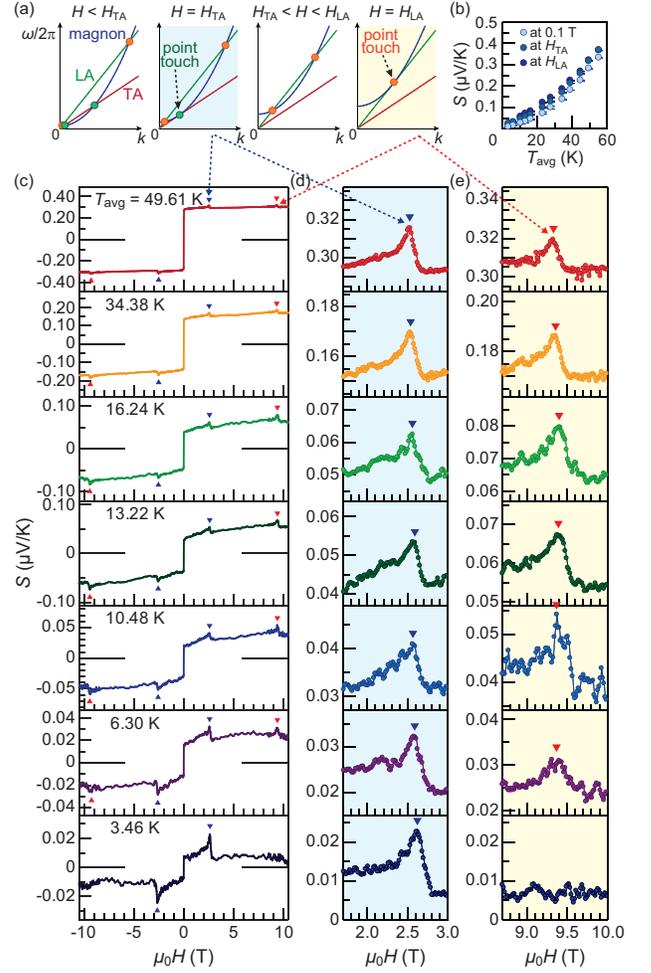}
\end{center}
\caption{(a) Magnon, TA-phonon, and LA-phonon dispersion relations for YIG
when $H<H_{\mathrm{TA}}$, $H=H_{\mathrm{TA}}$, $H_{\mathrm{TA}}<H<H_{\mathrm{%
LA}}$, and $H=H_{\mathrm{LA}}$. (b) $T_{\mathrm{avg}}$ dependence of the normalized SSE voltage $S$ at $H=0.1~\text{T}$, $H_{\mathrm{TA}}$, and $H_{%
\mathrm{LA}}$. (c) $S(H)$ of the Pt/YIG/%
GGG sample for various values of $T_{\mathrm{avg}}$ in the range of $%
|H|<10.5~\text{T}$. (d),(e) A blowup of $S(H)$ around (d) $H_{\mathrm{TA}}$
and (e) $H_{\mathrm{LA}}$.}
\label{fig:3}
\end{figure}

We carried out systematic measurements of the temperature dependence of the
SSE enhancement at $H_{\mathrm{TA}}$ and $H_{\mathrm{LA}}$. Figure \ref%
{fig:3}(c) shows the normalized SSE voltage $S\equiv (V/\Delta
T)(L_{T}/L_{V})$ as a function of $H$ for various average sample temperature 
$T_{\mathrm{avg}}~[\equiv (T_{\mathrm{H}}+T_{\mathrm{L}})/2]$. The amplitude
of the SSE signal monotonically decreases with decreasing $T$ in the present
temperature range \cite{SSE_Kikkawa2015PRB,SSE_Jin2015PRB} [see Fig. \ref%
{fig:3}(b)]. Importantly, the two peaks in $S$ at $H_{%
\mathrm{TA}}$ and $H_{\mathrm{LA}}$ exhibit different $T$ dependences [see
Figs. \ref{fig:3}(c), \ref{fig:3}(d), and \ref{fig:3}(e)]. The peak shape at 
$H_{\mathrm{TA}}$ becomes more prominent with decreasing $T$ and it is the
most outstanding at the lowest $T$. On the other hand, the $S$ peak at $H_{%
\mathrm{LA}}$ is suppressed below $\sim 10~\text{K}$ and it is almost
indistinguishable at the lowest $T$. This different $T$ dependence can be
attributed to the different energy scale of the branch crossing point for $%
H=H_{\mathrm{TA}}$ and $H=H_{\mathrm{LA}}$. The frequency of the magnon$-$%
LA-phonon intersection point is $0.53~\text{THz}=26~\text{K}$ ($\equiv
T_{\mathrm{MLA}}$), and it is more than 3 times larger than that of
the magnon$-$TA-phonon intersection point ($0.16~\text{THz}$). Therefore, for $T<T_{\mathrm{MLA}}
$, the excitation of magnons with energy around the magnon$-$LA-phonon
intersection point is rapidly suppressed, which leads to the disappearance of
the $S$ peak at $H_{\mathrm{LA}}$ at the lowest $T$. \par
The clear peak structures at low temperatures allow us to unravel the behavior
of the SSE around $H_{\mathrm{TA}}$ in detail. Increasing $H$ from small
values, $S$ increases up to a maximum
value at $H=H_{\mathrm{TA}}$, as shown in Fig. \ref{fig:3}(d)  ($T_{\mathrm{avg}}=3.46~\text{K}$). For fields
slightly larger than $H_{\mathrm{TA}},$ $S$ drops steeply to a value below
the initial one. The SSE intensity $S(i)$, where $i$ ($=0,1,2$) represents
the number of crossing points between the magnon and (TA-)phonon branch
curves [see also Fig. \ref{fig:2}(a)], can be ordered as $S(1)>S(2)>S(0)$ and
could be a measure of the number of magnon polarons.\par
%
%
%
%
%
%
%
%
%
%
%
The SSE is generated in three steps: (i) the temperature gradient excites
magnetization dynamics that (ii) at the interface to the metal becomes a 
particle spin current and (iii) is converted to a transverse voltage by
the ISHE. The latter two steps depend only weakly on the magnetic field. For
thick enough samples, the observed anomalies in the SSE originate from the
thermally excited spin current in the bulk of the ferromagnet. The
importance of the magnetoelastic coupling (MEC) for spin transport in magnetic
insulators has been established by spatiotemporally resolved pump-and-probe
optical spectroscopy \cite{Ogawa2015PNAS,Shen2015PRL}. Here we develop a
semiclassical model for the SSE in the strongly coupled magnon-phonon
transport regime \cite%
{Kamra2015PRB,Kittel1958PhysRev,Ruckriegel2014PRB,Shen2015PRL,Guerreiro2015PRB}. Our
model Hamiltonian consists of magnon ($\mathcal{H}_{\mathrm{mag}}$), phonon (%
$\mathcal{H}_{\mathrm{el}}$), and magnetoelastic coupling ($\mathcal{H}_{%
\mathrm{mec}}$) terms. In second-quantized form $\mathcal{H}_{\mathrm{mag}%
}=\sum_{{\mathbf{k}}}A_{\mathbf{k}}a_{{\mathbf{k}}}^{\dagger }a_{{\mathbf{k}}%
}+(B_{\mathbf{k}}/2)(a_{\mathbf{k}}^{\dagger }a_{-\mathbf{k}}^{\dagger }+a_{-%
\mathbf{k}}a_{\mathbf{k}})$, $\mathcal{H}_{\mathrm{el}}=\sum_{\mathbf{k},\mu
}\hbar \omega _{\mu \mathbf{k}}\left( c_{\mu {\mathbf{k}}}^{\dagger }c_{\mu {%
\mathbf{k}}}+\tfrac{1}{2}\right) $, and $\mathcal{H}_{\mathrm{mec}}=\hbar
nB_{\perp }(\frac{\gamma \hbar }{4M_{\mathrm{s}}\rho })^{1/2}\sum_{{\mathbf{k%
}},\mu }k\omega _{\mathbf{k}\mu }^{-1/2}e^{-i\phi }a_{\mathbf{k}}(c_{\mu -%
\mathbf{k}}+c_{\mu \mathbf{k}}^{\dagger })\times (-i\delta _{\mu 1}\cos
2\theta +i\delta _{\mu 2}\cos \theta -\delta _{\mu 3}\sin 2\theta )+h.c.$.
In spherical coordinates the wave vector ${\mathbf{k}}=k(\sin \theta \cos
\phi ,\sin \theta \sin \phi ,\cos \theta )$, $A_{\mathbf{k}}/\hbar =D_{%
\mathrm{ex}}k^{2}+\gamma \mu _{0}H+(\gamma \mu _{0}M_{\mathrm{s}%
}\sin ^{2}\theta)/2$, and $B_{\mathbf{k}}/\hbar = (\gamma \mu _{0}M_{\mathrm{s}%
}\sin ^{2}\theta)/2$. Here, $a_{\mathbf{k}}^{\dag }$ ($c_{\mu \mathbf{k}%
}^{\dag }$) and $a_{\mathbf{k}}$ ($c_{\mu \mathbf{k}}$) are magnon (phonon)
creation and annihilation operators, respectively. $B_{\perp }$ is the
magnetoelastic coupling constant, $\rho $ is the average mass density, $%
n=1/a_{0}^{3}$ is the number density of spins, and $a_{0}$ is the lattice
constant. The magnon dispersion from $\mathcal{H}_{\mathrm{mag}}$ is given
by Eq. (\ref{equ:magnon}), while the phonon dispersions are $\omega _{\mu 
\mathbf{k}}=c_{\mu }k$ with $\mu =1,2$ for the two transverse modes and $\mu
=3$ for the longitudinal one. $\delta_{\mu i}$ in $\mathcal{H}_{\mathrm{mec}}$ represents the Kronecker delta. By diagonalizing 
$\mathcal{H}_{\mathrm{mag}}+\mathcal{H}_{\mathrm{el}}+%
\mathcal{H}_{\mathrm{mec}}$ \cite{Colpa1978PhysicaA}, we obtain the
dispersion relation of the $i$-th magnon-polaron branch $\hbar \Omega _{i%
\mathbf{k}}$ and the corresponding amplitude $|{\psi }_{i{\mathbf{k}}%
}\rangle $. The magnon-polaron dispersions for $\theta =\pi /2$ and $\phi =0$
are illustrated in Figs.~\ref{fig:1}(c) and \ref{fig:1}(d), with a magnetic field $\mu
_{0}H=1.0$~T and $B_{\perp }/(2\pi )=1988$~GHz~\cite{Gurevich-Melkov_text}.\par
We assume diffuse transport that at low temperatures is limited by elastic
magnon and phonon impurity scattering~\cite{Ruckriegel2014PRB}. We employ
the Hamiltonian $\mathcal{H}_{\mathrm{imp}}=\sum_{\mu }\sum_{\mathbf{k},%
\mathbf{k}^{\prime }}c_{\mu \mathbf{k}}^{\dagger }v_{\mathbf{k},\mathbf{k}%
^{\prime }}^{\mathrm{ph}}c_{\mu \mathbf{k}^{\prime }}+\sum_{\mathbf{k},%
\mathbf{k}^{\prime }}a_{\mathbf{k}}^{\dagger }v_{\mathbf{k},\mathbf{k}%
^{\prime }}^{\mathrm{mag}}a_{\mathbf{k}^{\prime }}$, where, assuming $s$-wave
scattering, $v_{\mathbf{k},\mathbf{k}^{\prime }}^{\mathrm{ph}}=v^{\mathrm{ph}%
}$ and $v_{\mathbf{k},\mathbf{k}^{\prime }}^{\mathrm{mag}}=v^{\mathrm{mag}}$
denote the phonon and magnon impurity scattering potentials, respectively.
We compute the spin current driven by a temperature gradient \cite{SSE_Rezende2014PRB,SSE_Rezende2016JMMM} and thereby the
SSE in the relaxation-time approximation of the linearized Boltzmann
equation. The linear-response steady-state spin current ${\mathbf{J}}^{s}(%
\mathbf{r})=-\boldsymbol{\zeta }\cdot {\nabla }T$ is governed by the SSE
tensor $\boldsymbol{\zeta }:$ 
\begin{equation}
{\zeta }_{\alpha \beta }= \int \frac{d^{3}\mathbf{k}}{(2\pi )^{3}}%
\sum_{i}W_{i\mathbf{k}}^{s}\tau _{i\mathbf{k}}(\partial _{k_{\alpha }}\Omega
_{i\mathbf{k}})(\partial _{k_{\beta }}\Omega _{i\mathbf{k}})\partial _{T}f_{i%
\mathbf{k}}^{(0)}|_{T=T{(\mathbf{r})}}.  \label{Luqdef}
\end{equation}%
Here $W_{i\mathbf{k}}^{s}=|\langle 0|a_{\mathbf{k}}|\psi _{i\mathbf{k}%
}\rangle |^{2}$ is the intensity of the $i$-th magnon-polaron and $\tau _{i%
\mathbf{k}}$ is the relaxation time towards the equilibrium (Planck)
distribution function $f_{i{\mathbf{k}}}^{(0)}(\mathbf{r})=\left( \exp
\left( \hbar \Omega _{i\mathbf{k}}/\left( k_{\mathrm{B}}T(\mathbf{r})\right)
\right) -1\right) ^{-1}$. The relaxation time $\tau _{i\mathbf{k}}$ of the $%
i $-th magnon-polaron reads $\tau _{i\mathbf{k}}^{-1}=(2\pi /\hbar )\sum_{j%
\mathbf{k}^{\prime }}|\langle \psi _{j\mathbf{k}^{\prime }}|\mathcal{H}_{%
\mathrm{imp}}|\psi _{i\mathbf{k}}\rangle |^{2}\delta (\hbar \Omega _{i%
\mathbf{k}}-\hbar \Omega _{j\mathbf{k}^{\prime }})$. The strong-coupling
(weak scattering) approach is valid when $\tau _{i\mathbf{k}_{1,2}}^{-1}\ll
\Delta \Omega $, where $\Delta \Omega $ is the energy gap at the
anticrossing points $\mathbf{k}_{1,2}$. We disregard the Gilbert damping
that is very small in YIG.\par
\begin{figure}[tbh]
\begin{center}
\includegraphics{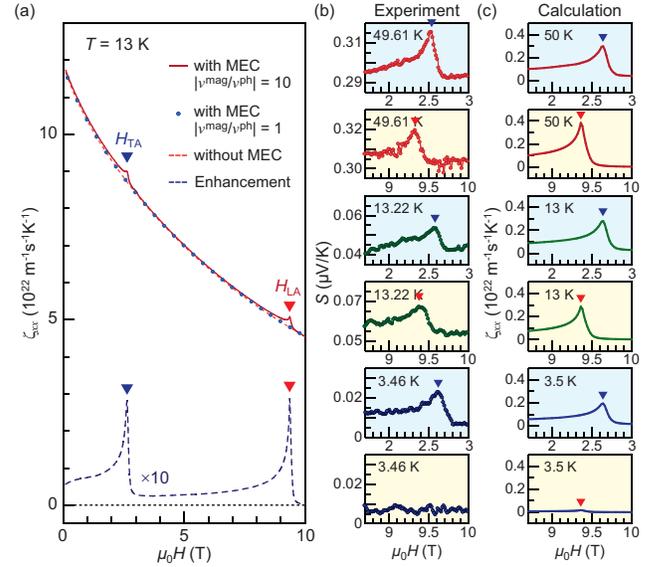}
\end{center}
\caption{(a) Calculated SSC $\protect\zeta _{xx}$ at $T=13~\text{K}$ as a
function of $H,$ with (red solid curve and blue circles) and without (red
dashed curve) magnetoelastic coupling (MEC). The red solid curve and the
blue circles are computed for ratios of the scattering potentials of $\left\vert
v^{\mathrm{mag}}/v^{\mathrm{ph}}\right\vert =10$ and $\left\vert v^{\mathrm{%
mag}}/v^{\mathrm{ph}}\right\vert =1$, respectively. The blue dashed curve
is a blowup of the difference between the red solid and dashed curves. (b) Experimental $S$
and (c) theoretical $\protect\zeta _{xx}$ after subtraction of the zero MEC results.}
\label{fig:4}
\end{figure}
%
%
From the experiments we infer the scattering parameters $|v^{\mathrm{mag}%
}|^{2}=10^{-5}$~s$^{-2}$ \cite{Boona2014PRB} and $\left\vert v^{\mathrm{mag}}/v^{\mathrm{ph}%
}\right\vert =10,$ i.e., the magnons are more strongly scattered than the phonons.
The computed longitudinal spin Seebeck coefficient (SSC) $\zeta _{xx}$ [Eq.~(%
\ref{Luqdef})] is plotted in Fig.~\ref{fig:4}(a). Switching on the
magnetoelastic coupling increases the SSC especially at the
\textquotedblleft touching\textquotedblright\ magnetic fields $H_{\mathrm{TA}%
}$ and $H_{\mathrm{LA}}$. At these points the group velocity of the magnon
is identical to the sound velocity. Nevertheless, spin transport can be
strongly modified when the ratio $\left\vert v^{\mathrm{mag}}/v^{\mathrm{ph}%
}\right\vert $ differs from unity. The SSC can be enhanced or suppressed
compared to its purely magnonic value. A high acoustic quality as implied by 
$\left\vert v^{\mathrm{mag}}/v^{\mathrm{ph}}\right\vert =10$ is beneficial
for spin transport and enhances the SSC by hybridization, as illustrated by
Fig.~\ref{fig:4}(a). When magnon and phonon scattering potentials would be
the same, i.e., $\left\vert v^{\mathrm{mag}}/v^{\mathrm{ph}}\right\vert =1,$
the anomalies vanish identically [see the blue circles in Fig.~\ref{fig:4}(a)].
The difference between the calculations with and without MEC agrees very well
with the peak features on top of the smooth background as observed in the
experiments, see Figs.~\ref{fig:4}(b) and \ref{fig:4}(c). We can rationalize the result
by the presence of a magnetic disorder that scatters magnons but not phonons.\par
%
%
%
%
%
%
%
%
%
%
%
%
%
%
%
Finally, we address the SSE background signal. The overall decrease of the
calculated $\zeta _{xx}$ is not related to the phonons, but reflects the
field-induced freeze-out of the magnons (that is suppressed in thin magnetic
films~\cite{SSE_Kikkawa2015PRB}). In the experiments, on the other hand,
the global $S$ below $\sim 30~\text{K}$ clearly increases with increasing $H$
[Fig. \ref{fig:3}(c)]. We tentatively attribute this discrepancy to an
additional spin current caused by the paramagnetic GGG substrate that, when
transmitted through the YIG layer, causes an additional voltage. Wu \textit{%
et al.} \cite{SSE_Wu2015PRL} found a paramagnetic SSE signal in a Pt%
/GGG sample proportional to the induced magnetization ($\sim $
a Brillouin function for spin 7/2) \cite{SSE_Wu2015PRL}. Indeed, the
increase of $S$ in the present Pt/YIG/GGG
sample is of the same order as the paramagnetic SSE in a Pt/%
GGG sample \cite{SSE_Kikkawa2015PRB}.\par
In conclusion, we observed two anomalous peak structures in the magnetic
field dependence of the spin Seebeck effect (SSE) in Pt/Y$%
_{3}$Fe$_{5}$O$_{12}$ (YIG) that appear at the onset of magnon-polaron
formation. The experimental results are well reproduced by a calculation in
which magnons and phonons are allowed to hybridize. Our results
show that the SSE can probe not only magnon dynamics but also phonon
dynamics. The magnitude and shape of the
anomalies contain unique information about the sample disorder, depending
sensitively on the relative scattering strengths of the magnons and phonons. \par
The authors thank S. Daimon, J. Lustikova, L. J. Cornelissen, and B. J. van Wees for valuable discussions. This work was supported by PRESTO \textquotedblleft Phase Interfaces for
Highly Efficient Energy Utilization\textquotedblright\ from JST, Japan,
Grant-in-Aid for Scientific Research on Innovative Area \textquotedblleft
Nano Spin Conversion Science\textquotedblright\ (No. 26103005, 
26103006), Grant-in-Aid for Scientific Research (A) (No. 15H02012, 25247056)
and (S) (No. 25220910) from MEXT, Japan, NEC Corporation, The Noguchi
Institute, the Dutch FOM Foundation, EU-FET Grant InSpin 612759, and DFG
Priority Programme 1538 \textquotedblleft Spin-Caloric Transport" (BA
2954/2). T.K. is supported by JSPS through a research fellowship for young
scientists (No. 15J08026).\par
%
%
%
%
%

%
%
%
\end{document}